# Theoretical approaches on alpha decay half-lives of the super heavy Nuclei

*S. S. Hosseini*[*1], H. Hassanabadi[1]*

[1]*Physics Department, Shahrood University of Technology, Shahrood, Iran*

[*]*Corresponding author, Tel.:+98 232 4222522; fax: +98 273 3335270
Email: seyedesamira.hosseini@gmail.com*

**Abstract**

We considered the systematics of α-decay (AD) half-life (HL) of super-heavy nuclei versus the decay energy and the total α-kinetic energy. We calculated the HL using the experimental $Q_\alpha$ values. The computed half-lives are compared with the experimental data and also with the existing empirical estimates and are found in good agreement. Also, we obtained α-preformation factors from the ratio between theoretical and experimental results for some super heavy nuclei SHN and evaluated the standard deviation. The results indicate the acceptability of the approach.

**Keywords.** Alpha decay, super heavy nuclei (SHN), half-life, preformation factor.

**PACS: No 23.70. +j; 23.60. +e**

## 1 Introduction

The first correlation of the empirical formula predicted by Geiger and Nuttall [1] and shaped the experimental values of $\log_{10}(T_{1/2})$ vs $Q_\alpha^{1/2}$. Independently, Gamow [2] and Gurney and Condon [3] analyzed the one-body problem for AD and derived the known Geiger-Nuttall (GN) correlation from first principles of quantum mechanics that formulated a function of the halftime, the energy $Q_\alpha$ and the proton number of daughter nucleus $Z_d$. Viola and Seaborg (VS) [4] considered the intercept parameters linear dependence on the charge number of the daughter nucleus by the work of Gallagher and Rasmussen [4, 5]. A linear relation between the Geiger-Nuttall law, $Z_d$ and $Q_\alpha$ quantity considered by Brown that the best representation was for describing the AD properties of super-heavy nuclei SHN. Royer (R) [6] suggested another empirical formula for AD HL that $\log(T_\alpha)$ depended on the decay energy, the atomic mass number and the charge number of the parent nucleus. Dong et.al [7] derive an expression of $Q_\alpha$ value based on the liquid drop model, which can be used as an input to quantitatively predict the half-lives of unknown nuclei. Alpha decay (AD) typically occurs in the heaviest nuclides. Alpha particles described in the investigations of radioactivity by Ernest Rutherford in 1899 [8] and Gamow had interpreted the theory of alpha decay (quantum penetration of α particles) via tunneling in 1928. The alpha particle is trapped in a potential well by the nucleus. There are many theoretical and experimental approaches which have investigated the AD, α cluster radioactivity models and the SHN presented in Ref. [9-27]. The first systematics of α-decay properties of super-heavy SHN was performed by studying the half-life versus kinetic energy (KE) correlations in terms of atomic number (Z) and mass number (A). The AD half-life (HL) obtained from clustering and scattering amplitudes given by self-consistent nuclear models for the nuclear shell structure and reaction dynamics for SHN with Z=104-120 were reported in Ref. [20]. Budaca and Silisteanu studied the AD of SHN within the framework of the shell-model rate theory also calculated the HLs and resonance scattering amplitudes with self-consistent models for the nuclear structure and reaction dynamics [28]. Silisteanu et al. solved the radial Schrödinger equation for coupled channels problem with outgoing asymptotic and resonance conditions for estimating the alpha-emission rates of ground and excited states of heaviest elements [29, 31]. Before time the nuclear shell model (NSM) predicted that the next magic proton number beyond Z = 82 would be Z = 114. Recently microscopic nuclear theories suggest a magic island around Z = 120, 124, or, 126 and N = 184. The heavy elements with Z = 107 - 112 have been successfully synthesized at GSI, Darmstadt and both theoretical and experimental facets of SHN are extensively discussed [32-34]. The life-times of several isotopes of heavy elements with Z = 102-120 calculated the quantum mechanical tunneling probability in a WKB framework and microscopic nucleus-nucleus potential with the DD (density dependent) M3Y effective nuclear interaction [35, 36]. This manuscript is organized as follows: Section 2 gives a brief description of the empirical approach to AD HL for isotopes of SHN. In Section 3 the penetration probability is summarized and evaluated the standard deviation. In Section 4 results and discussions are provided and the conclusion of the work is given as Section 5.

## 2. The Empirical approach

The Geiger–Nuttall (GN) law is given by $\log_{10} T_{1/2(\alpha)}^{GN} = aQ_\alpha^{-1/2} + b$ Where $a$ and b are the coefficients which are determined by fitting experimental data for each isotopic chain and $Q_\alpha$ (MeV) is the total energy of the α-decay process (α- decay Q value). The decay $Q_\alpha$ values for measured super-heavy α emitters can be obtained from the measured α-particle kinetic energy (KE) $E_\alpha$ using the following expression

$$Q_\alpha = \frac{A_p}{A_p - 4} E_\alpha^{\exp} + [6.53(Z_d - 2)^{7/5} - 8.0(Z_d - 2)^{2/5}].10^{-5} (MeV), \qquad (1)$$

where the first and second terms are the standard recoil and an electron shielding correction in a systematic manner, respectively as suggested by Perlman and Rasmussen [37] and A and Z are the mass and atomic numbers of the parent nucleus [38, 51], $E^{\exp}_\alpha$ is the measured kinetic energy of $\alpha$-particles, and the last term in Eq. (1) is the screening energy.

Dong et.al was proposed a formula for α-decay Q value of SHN based on a liquid drop model (LDM) [7, 39-41]. We have calculated the $Q_\alpha^{theor}$ value using the equation,

$$Q_\alpha^{theor} = \alpha Z A^{-4/3}(3A - Z) + \beta\left(\frac{N-Z}{A}\right)^2 + \gamma\left[\frac{|N-152|}{N} - \frac{|N-154|}{N-2}\right] + \delta\left[\frac{|Z-110|}{Z} - \frac{|N-112|}{Z-2}\right] + \varepsilon, \qquad (2)$$

Here Z, N and A are the proton, neutron and mass numbers of the parent nuclei, respectively. The first two terms are the contributions coming from the LDM Coulomb energy and symmetry energy, respectively, while the next two account for the neutron and proton shell effects of N = 152 and respectively Z = 110. The involved parameters were determined in [42-45] by fitting N=154 experimental $Q_\alpha$ data points and have the values α = 0.9373 MeV, β = −99.3027 MeV, γ = 16.0363 MeV, δ = −21.5983 MeV and ε = −27.4530 MeV [7]. Now we analyze three phenomenological formulas of the empirical formulas for half-life systematics of SHN. The first is the Royer (R) formula [6] which can be written as

$$\log_{10} T_{1/2(\alpha)}^R = aZ_p Q_\alpha^{-1/2} + bA_p^{1.6} Z^{1/2} + c. \qquad (3)$$

Where a, b and c are adjustable parameters that these coefficients referred to each ($Z_p$, $N_p$) parity of the parent nucleus combination that we illustrate with even–even (e–e), odd–even (o–e), even–odd (e–o) and odd–odd (o–o). These parameters represented in Ref. [6], we listed in table 1. The second one is the well-known Viola-Seaborg (VS) formula [4] which is written

$$\log_{10} T_{1/2(\alpha)}^{VS} = (aZ_p + b)Q_\alpha^{-1/2} + (cZ_p + d) + h_{Z-N}^{VS}, \qquad (4)$$

Where $Z_p$ is the charge number of the parent nucleus, $h_{Z-N}^{VS}$ is an even-odd hindrance term, and a, b, c and d are fitting parameters. The parameters used are, [46], also see table 1. The hindrance terms values obtained from the original paper of Viola and Seaborg. Other sets of parameters are constantly provided by fits on updated and new experimental data [47] or different sets of highly precise data [48].

**Table 1. The parameters taken from for the VS [4], R [6] and mB1 and mB2 [51] formulas.**

| | h | | a | | | | b | | | | c | | | | d |
|---|---|---|---|---|---|---|---|---|---|---|---|---|---|---|---|
| | VS | mB1 | VS | mB1 | R | mB2 | VS | mB1 | R | mB2 | VS | mB1 | R | mB2 | VS |
| e-e | - | - | 1.5744 | 13.0705 | 1.6672 | 10.8238 | -23.392 | 0.5182 | -1.2216 | 0.5966 | -0.2746 | -47.8867 | -26.3843 | -56.9785 | -33.9069 |
| e-o | 0.1.066 | 0.4666 | | | 1.4763 | 14.7747 | | | -1.3523 | 0.5021 | | | -15.8306 | -49.7080 | |
| o-e | 0.772 | 0.6001 | | | 1.1499 | 11.1462 | | | -1.0402 | 0.5110 | | | -12.6186 | -39.0096 | |
| o-o | 1.114 | 0.820 | | | 1.2451 | 14.7405 | | | -1.2134 | 0.4666 | | | -11.1310 | -41.7227 | |

The third one is the Brown formula, obtained from the semi-classical Wentzel- Kramers- Brillouin (WKB) approximation and fit to the experimental data [49, 50] which is given by

$$\log T_\alpha^B = 9.54 Z_d^{0.6} Q_\alpha^{-1/2} - 51.37, \qquad (5)$$

The parameters are determined by fitting to the available experimental data from [28]. Budaca et al. expressed the modified Brown formula with comparison and fitting of the VS and R formulas. The first modified Brown fit (mB1) will have the parameters a, b and c parity independent with an additional hindrance term differentiated by parity [51]

$$\log T_\alpha^{mB1} = a(Z_p - 2)^b Q_\alpha^{-1/2} + c + h_{Z-N}^{mB1}, \qquad (6)$$

The parameters a, b and c parity and hindrance terms are for mB1 formula. For the modified Brown formula (mB2) in work Ref. [51] is chosen as:

$$\log T_\alpha^{mB2} = a_{Z-N}(Z_p - 2)^{b_{Z-N}} Q_\alpha^{-1/2} + c_{Z-N}, \qquad (7)$$

The $a$, b and c parameters have shown in table 1. The parameters of the empirical formulae are usually determined by fitting to a large amount of data which on the other hand may yield notable errors. This means that the relation between half-live, reaction energy and number of constituent nucleons is in fact quite complicated [51].

### 3. α-Preformation factor

Lovas et.al discussed about microscopic theory of alpha cluster radioactivity decay in 1998. The preformation probability defined in the quantum mechanical for two-cluster component in the bound initial state of the parent nucleus [52]. It describes the influences of the different nuclear structure of properties of the parent for instance their isospin asymmetry of the even-even nuclei [53], shell closure from ground and isomeric states and pairing effects [54, 55] and a double folding procedure using M3Y plus Coulomb two-body forces of the quadrupole deformations [56, 57]. Furthermore, several theoretical and experimental efforts have been made to calculate the preformation factor $S_\alpha$ [58-61]. The preformation factor $S_\alpha$ is obtained from the ratio of the calculated and the experimental half-lives. Also, the $S_\alpha$ can be used for the prediction of half-lives of unknown super-heavy nuclei in a consistent way. The preformation factor may be also obtained from work of Mohr [62] reporting $S_\alpha = T_{1/2(\alpha)}^{cal}(s) / T_{1/2(\alpha)}^{exp}(s)$. We plotted the ratio $T_{1/2(\alpha)}^{cal} / T_{1/2(\alpha)}^{exp}$ versus the neutron number of the daughter nucleus ($N_d$). $\log_{10}(T_{1/2(\alpha)}^{exp}(s))$ values are reported in Table (2).

**Table 2. logarithm α decay half-lives for SHN with various theoretical estimations and comparison with the results obtained by VS, R, the two versions of mB empirical formulas and experimental data.**

| $A_p$ | $Z_p$ | $N_p$ | $Q_\alpha$(MeV)[23] | $Q^{theor}$(MeV)[7] | $E_\alpha$(MeV)[23] | $\log(T_{1/2\alpha}(s))$ | | | | |
|---|---|---|---|---|---|---|---|---|---|---|
| | | | | | | exp | VS | R | mB1 | mB2 |
| $^{267}$Rf | 104 | 163 | ≤8.22 | 7.85 | - | 3.9180 | 3.0295 | 2.3343 | 2.3154 | 2.4781 |
| $^{271}$Sg | 106 | 165 | 8.65±0.08 | 8.50 | 8.53±0.08 | 2.1583 | 2.5463 | 1.8603 | 1.7543 | 1.8731 |
| $^{275}$Hs | 108 | 167 | 9.44±0.07 | 9.15 | 9.30±0.07 | 0.8239 | 0.8313 | 0.1576 | 0.1537 | 0.1789 |
| $^{279}$Ds | 110 | 169 | 9.84±0.06 | 9.80 | 9.70±0.06 | -0.7447 | 0.5061 | -0.1485 | -0.2319 | -0.2403 |
| $^{281}$Ds | 110 | 170 | ≤9.05 | 9.65 | 9.00527 | 0.9822 | 1.3765 | 0.0869 | 0.9449 | 1.3941 |
| $^{282}$Cn | 112 | 170 | ≤10.82 | 9.798 | 10.7741 | -3.3010 | -2.9445 | -4.1132 | -2.8041 | -3.0098 |
| $^{283}$Cn | 112 | 171 | 9.67±0.06 | 9.65 | 9.54±0.06 | 0.6020 | 1.4590 | 0.8263 | 0.4687 | 0.4793 |
| $^{284}$Cn | 112 | 172 | ≤9.85 | 9.50 | 9.804097 | -0.9956 | -0.3740 | -1.6237 | -0.6427 | -0.4225 |
| $^{285}$Cn | 112 | 173 | 9.29±0.06 | 9.36 | 9.16±0.06 | 1.5314 | 2.6255 | 1.9576 | 1.4495 | 1.5072 |
| $^{286}$Fl | 114 | 172 | 10.35±0.06 | 10.45 | 10.20±0.06 | -0.7958 | -0.8446 | -2.0527 | -1.1306 | -0.9273 |
| $^{287}$Fl | 114 | 173 | 10.16±0.06 | 10.30 | 10.02±0.06 | -0.2924 | -0.2847 | -0.2466 | 0.1155 | 0.7222 |
| $^{288}$Fl | 114 | 174 | 10.09±0.07 | 10.16 | 9.95±0.07 | -0.0969 | -0.1444 | -1.4020 | -0.5472 | -0.2280 |
| $^{289}$Fl | 114 | 175 | 9.96±0.06 | 10.02 | 9.82±0.06 | 0.4313 | 1.2960 | 0.6539 | 0.2313 | 0.2160 |
| $^{290}$Lv | 116 | 174 | 11.00±0.08 | 11.08 | 10.85±0.08 | -1.8239 | -1.9466 | -3.1346 | -2.1267 | -2.0454 |
| $^{291}$Lv | 116 | 175 | 10.89±0.07 | 10.95 | 10.74±0.07 | -2.2006 | -0.5974 | -1.1729 | -1.4263 | -1.5344 |
| $^{292}$Lv | 116 | 176 | 10.80±0.07 | 10.80 | 10.66±0.07 | -1.7447 | -1.4544 | -2.6886 | -1.7204 | -1.5575 |
| $^{293}$Lv | 116 | 177 | 10.67±0.06 | 10.67 | 10.53±0.06 | -1.2757 | -0.0445 | -0.6556 | -0.9699 | -1.0563 |
| $^{294}$Og | 118 | 176 | 11.81±0.06 | 11.71 | 11.65±0.06 | -2.744 | -3.3125 | -4.4707 | -3.3173 | -3.4016 |

We presented this factor for SHN in Table (2). The straight line depicts a good fit to the preformation factor versus the neutron number of nucleus. For example, we obtained the $S_\alpha^{mB2}$ value 1.00 for $^{293}$Cn. The value for $S_\alpha^{VS}$, $S_\alpha^{R}$ and $S_\alpha^{mB1}$ are obtained as 0.1, 0.58 and 0.83, respectively. For judging the agreement between the experimental and calculated values, we have evaluated the standard deviation, σ, for the α-decay half-lives. The standard deviation is given by [63],

$$\sigma = \sqrt{\frac{1}{N-1}\sum_{i=1}^{N}\left[\log_{10} T_{(1/2\alpha)i}^{theor} - \log_{10} T_{(1/2\alpha)i}^{\exp}\right]^2}. \qquad (8)$$

The results are summarized in Table 3.

Table. 3. The standard deviation obtained with the Royer formula with the values of parameters taken from [10].

| σ (Eq. (8)) | | | |
|---|---|---|---|
| VS | R | mB1 | mB2 |
| 0.7808 | 0.9655 | 0.5675 | 0.5709 |

**4 Results and discussions**

We used two fitting schemes with the well-known empirical correlations Viola-Seaborg (VS) and Royer (R) formula and compared the results with the two modified versions of the Brown (B) formula; mB1 and mB2 [51]. In Table 2 we have calculated the half-lives for some super-heavy nuclei. The first, second and third columns represent the mass, proton and neutron numbers of the parent. The fourth, fifth and sixth column is the decay energy ($Q_\alpha$) in MeV from Eq. (1) taken from [23], the theoretical decay energy from Eq. (2) [7, 39-41] and the alpha kinetic energy ($E_\alpha$) in work Oganessian et.al. [23] and listed the results in Table 2. The seventh columns is the experimental and the calculated half-lives with VS [4], R [6], mB1 and mB2 [51], respectively. We calculated α-decay half-lives by comparing with the empirical formula for the SHN. For example, the half-life $^{267}$Rf of VS value is 3.02180 that is better than R = 2.3343, mB1=2.3154 and mB2=2.4781 but for $^{285}$Cn of mB2 value is 1.5072 that is better than VS= 2.5463, R= 1.8603 and mB1= 1.4495. In Fig. 1 we plotted the preformation factor ($S_\alpha$) for VS, V, mB1 and mB2 vs. versus neutron number of the daughter.

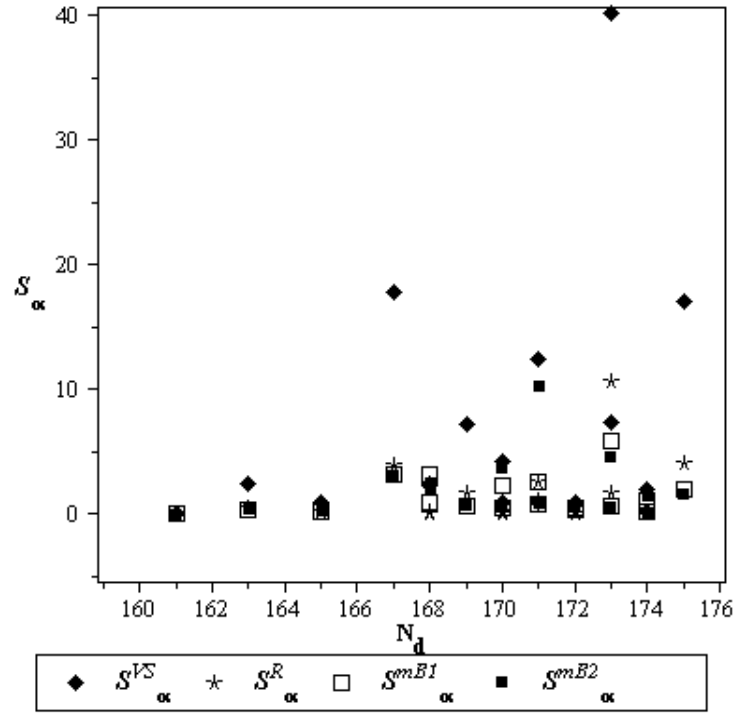

**Fig. 1.** $T^{x}_{1/2(\alpha)} / T^{\exp}_{1/2(\alpha)}$ **(The preformation factors $S^{x}_{\alpha}$ for several super heavy α-emitter) versus the neutron number of the daughter nucleus ($N_d$) are shown for comparison.**

In Fig. 2 we have shown how log $T_{1/2}$ (s) increases when decay energy increases. The behavior is in complete agreement with the Geiger–Nuttall rule.

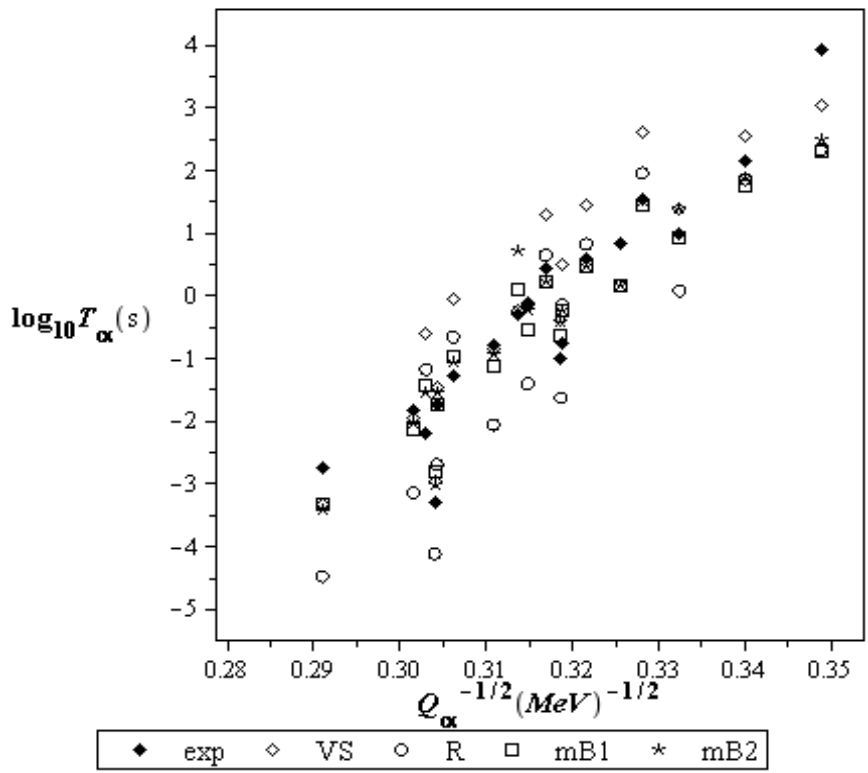

**Fig. 2. Calculated $\log_{10} T_{1/2(\alpha)}$ are plotted versus the effective decay energy $Q^{-1/2}_{(\alpha)} (MeV)^{-1/2}$.**

In Fig. 3 we showed separately for the $S_\alpha$ versus $N_d$ that VS values is better than R, mB1 and mB2 for some SHN.

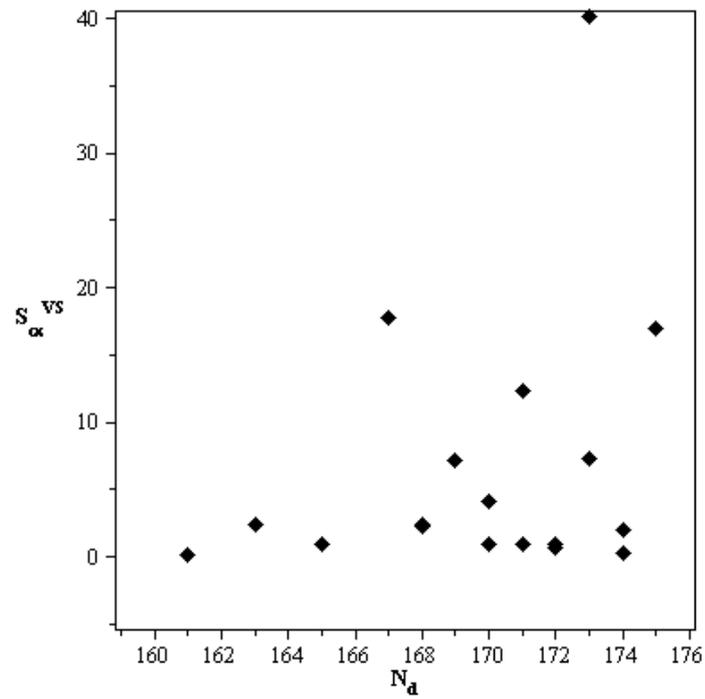

**(3-a) the preformation factor for VS**

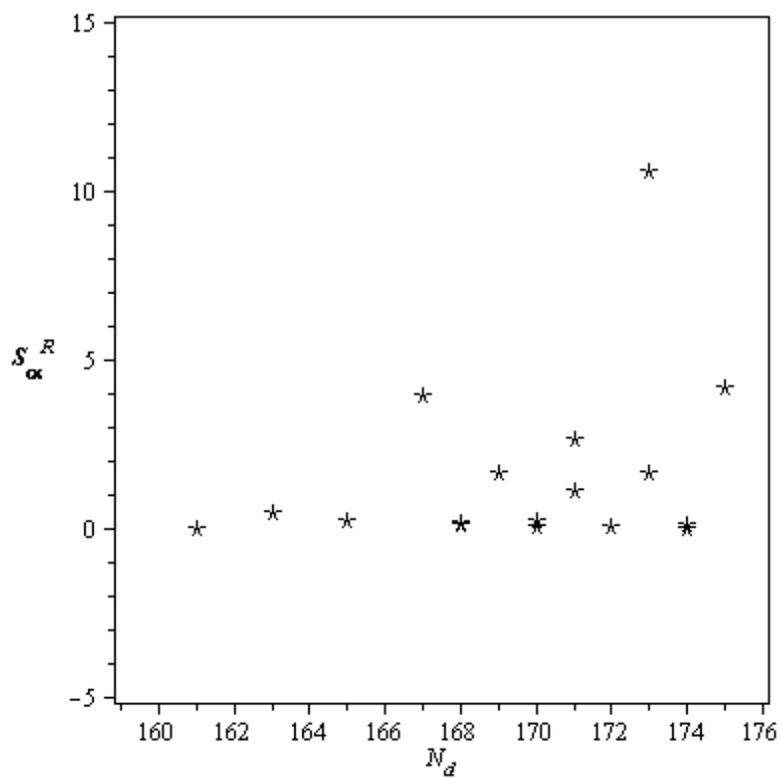

**(3-b) the preformation factor for R**

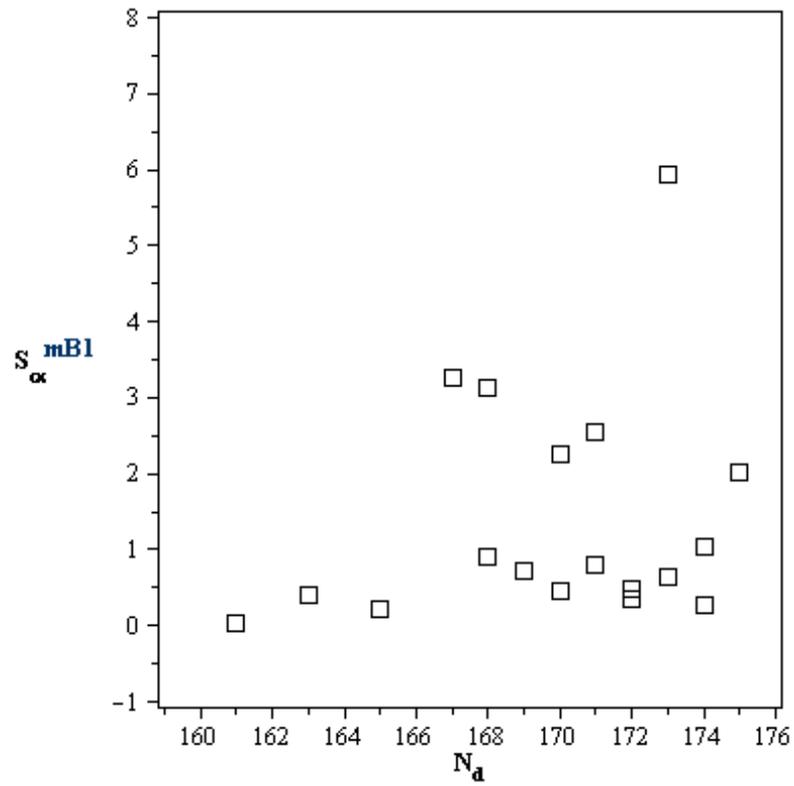

**(3-c) the preformation factor for mB1**

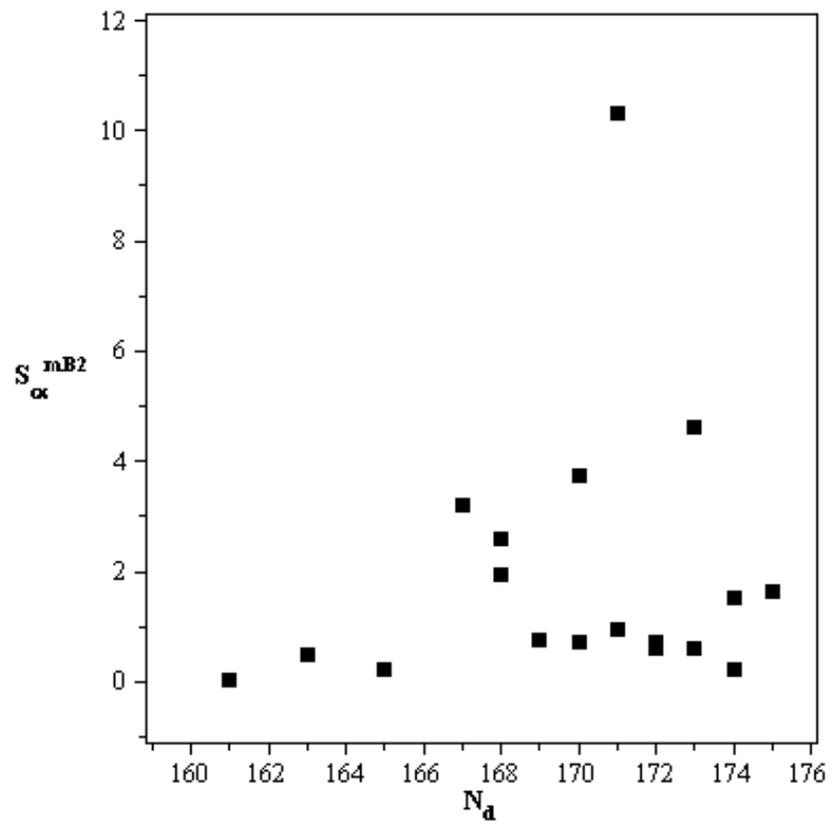

**(3-d) the preformation factor for mB2**

**Fig. 3.** The preformation factors $S_\alpha^x$ for SHN α-emitter versus the neutron number of the daughter nucleus ($N_d$).

In Fig. 4. Calculated $\log_{10} T_{1/2(\alpha)}$ s are plotted versus the effective decay energy $Q_\alpha^{-1/2}$ (MeV$^{-1/2}$) for Ds, Cn, Fl and Lv which shows the increasing behavior of log T$_{1/2}$ for increasing the effective decay energy. We have also compared the experimental and calculation data in Fig. 4. The results show an acceptable agreement with the experimental data. Indeed, the trend depicted in Fig. 4 for of Ds, Cn, Fl and Lv does indicate a suitable correlation between the half-life and the α-energy available for decay, resembling a Geiger–Nuttall-like law.

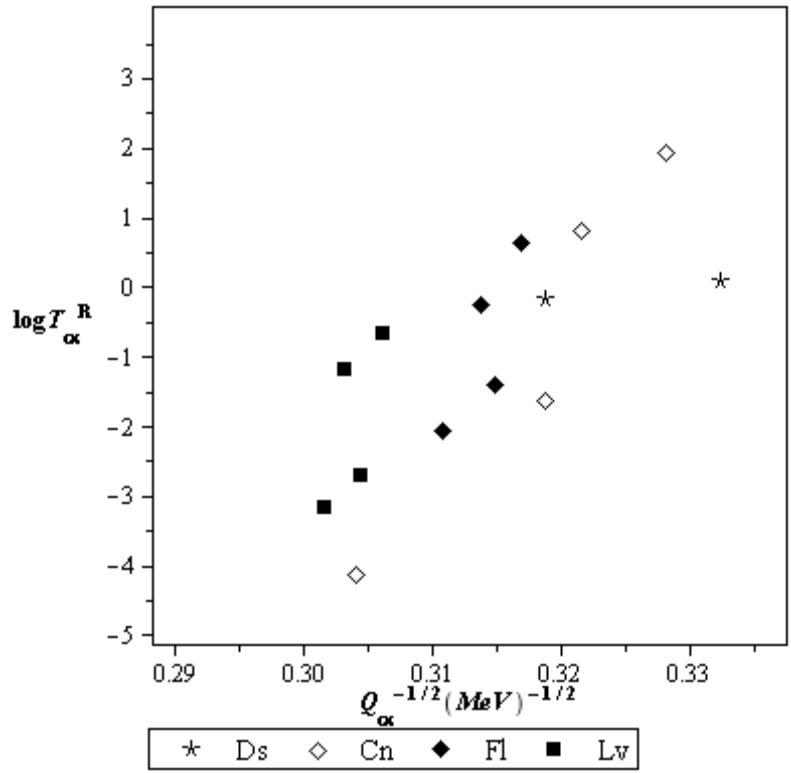

(4-a) for R

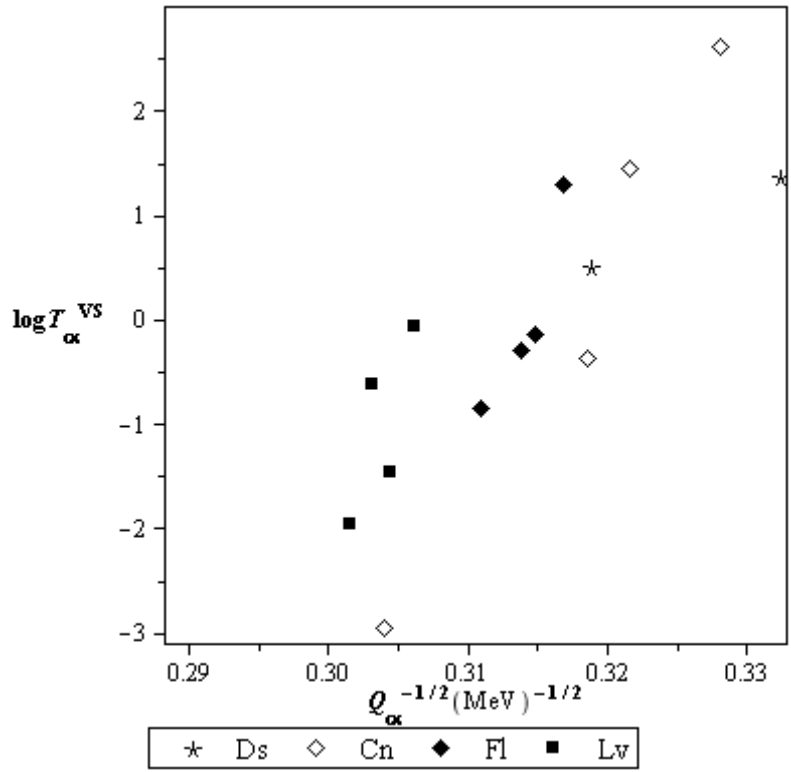

**(4-b) for VS**

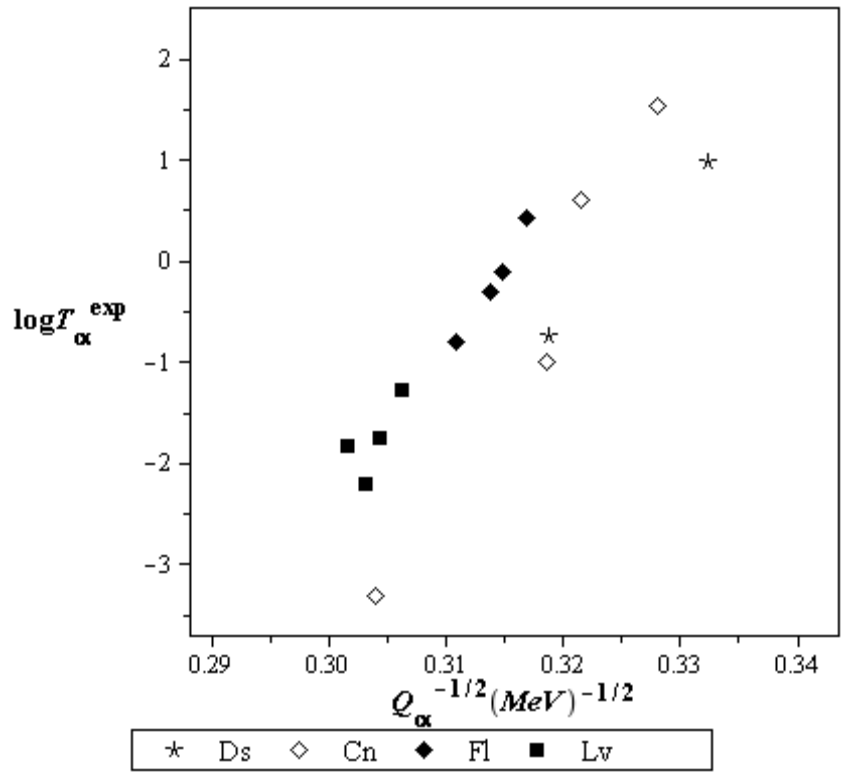

**(4-c) for experimental**

**Fig. 4.** Calculated $\log_{10} T_{1/2(\alpha)}$ are plotted versus the effective decay energy $Q_{(\alpha)}^{-1/2}$ for $Z_p$=110, 112, 114, 116.

# 5 Conclusion

In this manuscript, we considered the alpha decay half- lives for some super heavy nuclei as, Rf, Sg, Hs, Ds, Cn, Fl, Lv and Og and analyzed using the Viola-Seaborg and Royer formulae and a new analysis in the Brown formula [51]. The computed half-life values are compared with the experimental data and indicate acceptable agreement with some of the systematic of empirical correlation. From the ratio of the calculated and the experimental half-life, plotted versus $N_d$, a preformation factor for alpha decay is deduced. We depicted some results using the empirical and theoretical ways by comparison with experimental data for SHN. Finally, we calculated the standard deviation of the logarithm of half-life and the comparison models depicted in table 3 thus found to be 0.7808, 0.9655, 0.5675 and 0.5709 for VS, R, mB1 and mB2, respectively.

# Acknowledgments

It is a pleasure for authors to thank the kind referees for his many useful comments on the manuscript.